\documentclass{elsart}

\usepackage[mathlines]{lineno}
\usepackage{amssymb,amsmath,verbatim,graphicx}

\newcommand\R{\mathbb R}
\newcommand\C{\mathbb C}

\newcommand\diag{\operatorname{diag}}
\newcommand\im{\operatorname{Im}}
\newcommand\pr{\boldsymbol{\pi}_{GM}}
\newcommand\pI{\boldsymbol{\pi}_I}

\begin{document}

\begin{frontmatter}

\title{Identifying evolutionary trees and substitution parameters
for the general Markov model with invariable sites}

\thanks{This research was supported in part by the Institute for
  Mathematics and its Applications, with funds provided by the National
  Science Foundation.  We thank the IMA for its hospitality.}

\author{Elizabeth S. Allman},
\ead{e.allman@uaf.edu}
\author{John A. Rhodes\corauthref{cor}}
\ead{j.rhodes@uaf.edu}
\corauth[cor]{Corresponding author.}

\address{Department of Mathematics and Statistics\\University of
  Alaska Fairbanks\\PO Box 756660\\Fairbanks, AK 99775}

\date{February 22, 2007}

\begin{abstract}
  The general Markov plus invariable sites (GM+I) model of biological
  sequence evolution is a two-class model in which an unknown
  proportion of sites are not allowed to change, while the remainder
  undergo substitutions according to a Markov process on a tree. For
  statistical use it is important to know if the model is
  identifiable; can both the tree topology and the numerical
  parameters be determined from a joint distribution describing
  sequences only at the leaves of the tree?  We establish that for
  generic parameters both the tree and all numerical parameter values
  can be recovered, up to clearly understood issues of `label
  swapping.'  The method of analysis is algebraic, using phylogenetic
  invariants to study the variety defined by the model.  Simple
  rational formulas, expressed in terms of determinantal ratios, are
  found for recovering numerical parameters describing the invariable
  sites.

\end{abstract}

\end{frontmatter}

\section{Introduction}\label{sec:intro}

If a model of biological sequence evolution is to be used for
phylogenetic inference, it is essential that the model parameters of
interest --- certainly the tree parameter and usually the numerical
parameters
--- be identifiable from the joint distribution of states at the
leaves of the tree. Though often unstated, the assumption that model
parameters are identifiable underlies the use of both Maximum
Likelihood and Bayesian inference methods. As increasingly
complicated models, incorporating across-site rate variation,
covarion structure, or other types of mixtures, are implemented in
software packages, there is a real possibility that
non-identifiability could confound data analysis. Unfortunately, our
theoretical understanding of this issue lags well behind current
phylogenetic practice.

One natural approach to proving the identifiability of the tree
topology relies on the definition of a phylogenetic distance for the
model, and the $4$-point condition of Buneman \cite{Bun}. For
instance, Steel \cite{S94} used the log-det distance to establish
the identifiability of the tree topology under the general Markov
model and its submodels. Such a distance-based argument shows
additionally that $2$-marginalizations of the full joint
distribution suffice to recover the tree parameter, since distances
require only two-sequence comparisons. Once the tree has been
identified, the numerical parameters giving rise to a joint
distribution for the general Markov model
can be determined by an argument of Chang
\cite{MR97k:92011}.

However, for more general mixture models and rates-across-sites
models no appropriate definition of a distance is known, so proving the
identifiability of the tree parameter requires a different approach.
(Though distance measures have been developed for GTR models with
rate-substitution \cite{GuLi96,WadSt97}, these require that one know
the rate distribution completely, and identifiability of the rate distribution
has yet to be addressed.
Although identifiability of the popular GTR+I+$\Gamma$ model of
sequence evolution was considered in \cite{Rog01}, there are gaps in
the argument, as was pointed out to us by An\'e \cite{AnePC}.)

In \cite{ARidtree}, the viewpoint of algebraic geometry is used to
show the generic identifiability of the tree parameter for the
covarion model of \cite{MR1604518} and for certain mixture models with
a small number of classes.
Though this result is far more general than previous identifiability results,
it still fails to cover the type of rate-variation models
currently in common use for data analysis, and does not
address identifiability of numerical parameters at all.
Much more study of the identifiability question is needed.

\smallskip

In this paper, we focus on the \emph{general Markov plus invariable
 sites}, GM+I,  model of sequence evolution, a model that encompasses
the GTR+I model that is of more immediate interest to practitioners.
Note that previous work on GM+I by Baake \cite{MR1664261}
focused on \emph{non}-identifiability. In that paper
parameter choices for the $2$-state GM+I model
on two distinct 4-taxon trees
are constructed  that give rise to the same pairwise
joint distributions ($2$-marginals).  As both sets of parameters have
50\% invariable sites, this shows that the
identifiability of the tree parameter cannot generally hold on the basis of
2-sequence comparisons, even if the distribution of rate factors is
known. Furthermore, it shows a well-behaved phylogenetic distance cannot
be defined for this model, as existence of such a distance would imply tree
identifiability.

Here we prove that all parameters for the GM+I model are indeed
identifiable, through $4$-sequence comparisons.  By identifiable, we
mean \emph{generically identifiable} in a geometric sense: For a
fixed tree, the set of numerical parameters for which the joint
distribution could have arisen from either a) a different tree, or
b) a `significantly different' (in a sense to be made clear later)
choice of numerical parameters on the same tree, is of strictly
lower dimension than that of the full numerical parameter space.
(For a concrete example of generic identifiability,
recall the results of Steel and Chang on the general Markov model:
assumptions that
the Markov edge matrices $M_e$ have determinant $\ne 0,1$ and that the
distribution of states at the root has strictly positive entries ensure
identifiability of all parameters. These are
generic conditions.) Thus for natural probability distributions on
the parameter space, with probability one a choice of parameters is
generic.

\medskip

Although identifiability of the tree parameter for GM+I follows from
more general results in \cite{ARidtree}, that paper did not consider
identifiability of numerical parameters. Our arguments here are
tailored to GM+I and yield stronger results
addressing numerical parameters as well as the tree. Our approach
is again based on the determination of \emph{phylogenetic
invariants} for the model. While the invariants described in
\cite{ARidtree} are invariants for more general models than GM+I,
the ones given in this paper apply only to GM+I and its submodels,
and are of much lower degree. As a byproduct of the development of
these GM+I invariants, we are led to rational formulas for
recovering all the parameters related to the invariable sites from a
joint distribution.  Indeed, these formulas are crucial to our
identification of numerical parameters.

These formulas can be viewed as GM+I analogs of the formulas for the
proportion of invariable sites in group-based+I models that were
found by the capture-recapture argument of \cite{SHL00}.  In the
group-based setting, those formulas were developed into a heuristic
means of estimating the proportion of invariable sites from data
without performing a full tree inference. This has been implemented
in {\tt SplitsTree4} \cite{splitsTree4}. However, it remains unclear
whether a similar useful heuristic can be found for the formulas
presented in this paper.

\smallskip

Since our algebraic methods at times employ computational commutative
algebra software packages, and these tool are not commonly used
in the phylogenetics literature, we have included some examples of
code in Appendix \ref{app:code}.

\section{The GM+I Model}\label{sec:gmi}

Let $T$ denote an\emph{ $n$-taxon tree}, by which we mean a tree
with $n$ leaves labeled by the taxa $a_1,a_2,\dots, a_n$ and all
internal vertices of valence at least 3. We say $T$ is \emph{binary}
if all internal nodes have valence exactly 3.

We begin by describing the parameterization of the $\kappa$-state GM+I
model of sequence evolution along $T$, where $\kappa=4$ corresponds to
usual models of DNA evolution.  The \emph{class size parameter}
$\delta$ denotes the probability that any particular site in a
sequence is invariable: conceptually, the flip of a biased coin
weighted by $\delta$ determines if a site is allowed to undergo state
transitions. If a site is invariable, it is assigned state
$i\in[\kappa]=\{1,2,\dots,\kappa\}$ with probability $\pi_I(i)$. Here
$\boldsymbol \pi_I=(\pi_I(1),\dots,\pi_I(k))$ is a vector of
non-negative numbers summing to 1 giving the state distribution for
invariable sites.

All sites that are not invariable mutate according to a common set
of parameters for the GM model, though independently of one another.
For these sites, we associate to each node (including leaves) of $T$
a random variable with state space $[\kappa]$.  Choosing any node
$r$ of $T$ to serve as a root, and directing all edges away from
$r$, let $T_r$ denote the resulting directed tree $T$.  A \emph{root
distribution} vector $\pr= (\pi_{GM}(1),\dots, \pi_{GM}(\kappa))$,
with non-negative entries summing to $1$, has entries $\pr(j)$
specifying the probability that the root variable is in state $j$.
For each directed edge $e = (v \to w)$ of $T_r$, let $M_e$ be a
$\kappa\times \kappa$ Markov matrix, so that $M_e(i,j)$ specifies
the conditional probability that the variable at $w$ is in state $j$
given that the variable at $v$ is in state $i$. Thus entries of all
$M_e$ are non-negative, with rows summing to 1.

For the GM+I model on an $n$-taxon tree $T$ with edge set $E$, the
stochastic parameter space $S \subset {[0,1]}^N$
is of dimension
$N =1 + (\kappa-1) +(\kappa - 1) + |E|\kappa(\kappa - 1) = 2\kappa - 1 +
|E|\kappa(\kappa - 1).$
The parameterization map giving the joint
distribution of the variables at the leaves of $T$ is
denoted by
\begin{linenomath}
\begin{align*}
  \phi_T: S &\longrightarrow {[0,1]}^{\kappa^n},\\
  \mathbf{s} &\longmapsto P.
\end{align*}
\end{linenomath}
We view $P$ as an $n$-dimensional $\kappa \times \dots \times
\kappa$ array, with dimensions corresponding to the ordered taxa
$a_1,a_2,\dots,a_n$, and with entries indexed by the states at the
leaves of $T$. The entries of $P$ are polynomial functions in the
parameters $\mathbf{s}$ explicitly given by
\begin{linenomath}
\begin{multline}
  P(i_1, \dots, i_n) =\\
  \delta\, \epsilon (i_1,i_2,\dots i_n) \pI(i_1) +(1-\delta)
  \sum_{(j_v) \in \mathcal{H}} \left( \pr(j_r) \prod_{e} M_e(j_{v_i},
    j_{v_f})\right).\label{eq:Pdef}
\end{multline}
\end{linenomath}
Here $\epsilon(i_1,i_2,\dots i_n)$ is 1 if all $i_j$ are equal and 0
otherwise, the product is taken over all edges $e=(v_i\to v_f)\in E$,
and the sum is taken over the set of all possible assignments of
states to nodes of $T$ extending the assignment $(i_1, \dots, i_n)$ to
the leaves: If $V$ is the set of vertices of $T$ then
\begin{linenomath}
$$\mathcal{H} = \left\{(j_v) \in [\kappa]^{|V|} \mid j_v = i_k \mbox{
  if } v \mbox{ is a leaf labeled by $a_k$} \right\}.
$$
\end{linenomath}
For notational ease, the entries of $P$, the \emph{pattern
  frequencies}, are also denoted by $p_{i_1 \dots i_n} = P(i_1, \dots,
i_n)$.

We note that while a root $r$ was chosen for the tree in order to
explicitly describe the GM portion of the parameterization of our
model, the particular choice of $r$ is not important. Under mild
additional restrictions on model parameters, changing the root
location corresponds to a simple invertible change of variables in
the parameterization. (See \cite{SSH94}, \cite{AR03}, or \cite{ARgm}
for details.) This justifies our slight abuse of language in
referring to the GM or GM+I model on $T$, rather than on $T_r$, and
we omit future references to root location.

Note that equation (\ref{eq:Pdef}) allows us to more succinctly
describe any $P\in \im(\phi_T)$ as
\begin{linenomath}
\begin{equation}P= (1-\delta) P_{GM} + \delta P_I
\label{eq:decomp}\end{equation}
\end{linenomath}
where $P_{GM}$ is an array in the
image of the GM parameterization map on $T$ and
$P_I=\diag(\boldsymbol \pi_I)$ is an $n$-dimensional array whose
off-diagonal entries are zeros and whose diagonal entries are those
of $\pi_I$.

\section{Model Identifiability}\label{sec:modelId}

We now make precise the various concepts of identifiability of a
phylogenetic model. To adapt standard statistical language to the
phylogenetic setting, for a fixed set $A$ of $n$ taxa and $\kappa\ge
2$, consider a collection $\mathcal M$ of pairs $(T,\phi_T )$, where
$T$ is an $n$-taxon tree with leaf labels $A$, and $\phi_T:S_T\to
[0,1]^{\kappa^n}$ is a parameterization map of the joint
distribution of pattern frequencies for the model on $T$.  We say
\emph{the tree parameter is identifiable} for $\mathcal M$ if for
every $P\in \cup_{(T,\phi_T)\in
  \mathcal M} \im(\phi_T)$, there is a unique $T$ such that $P\in
\im(\phi_T)$. We say that \emph{numerical parameters are identifiable on a
tree $T$} if the map $\phi_T$ is injective, that is if for
every $P\in\im(\phi_T)$ there is a unique $\mathbf s\in S_T$ with
$\phi_T(\mathbf s)=P$. We say the \emph{model $\mathcal M$ is identifiable}
if the tree parameter is identifiable, and for each tree the numerical
parameters are identifiable.

It is well-known that such a definition of identifiability is too
stringent for phylogenetics. First, unless one restricts parameter
spaces, there is little hope that the tree parameter be identifiable:
One need only think of any standard model on a binary 4-taxon tree in
which the Markov matrix parameter on the internal edge is the identity
matrix. Any joint distribution arising from such a parameter choice
could have as well arisen from any other 4-taxon tree topology.

Even if such `special' parameter choices are excluded so the tree
parameter becomes identifiable, identifiability of numerical
parameters also poses problems, as noted  by Chang
\cite{MR97k:92011}. For example, consider the 3-taxon tree with the
GM model. Then multiple parameter choices give rise to the same
joint distribution since the labeling of the states at the internal
node can be permuted in $\kappa!$ ways, as long as the Markov matrix
parameters are adjusted accordingly \cite{AR03}. The occurrence of
this sort of `label-swapping' non-identifiability in statistical
models with hidden (unobserved) variables is well-known, but is not
of great concern. However, even for this model more subtle forms of
non-identifiability can occur, in which infinitely many parameter
choices lead to the same joint distribution. These arise from
singularities in the model, and can be avoided by again restricting
parameter space. Such `generic' conditions for the GM model have
already been mentioned in the introduction.

We therefore refine our notions of identifiability. Because we are
concerned primarily with model where the maps $\phi_T$ are given by
polynomials, we give a formulation appropriate to that setting.
Recall that given any collection $\mathcal F$ of polynomials
in $N$ variables, their common zero set,
\begin{linenomath}
$$
V(\mathcal F)=\{z\in \C^N \mid f(z)=0 \text{ for all } f\in \mathcal F\},
$$
\end{linenomath}
is the \emph{algebraic variety} defined by $\mathcal F$. If the algebraic
variety is a proper subset of $\C^N$, then it is said to be \emph{proper}.

\begin{defn}
Let $\mathcal M$ be a model on a collection of $n$-taxon trees, as
defined above.
\begin{enumerate}
\item  We say \emph{the tree parameter is generically identifiable}
for $\mathcal M$ if for each tree $T$ there exists a proper
algebraic variety $X_T$ with the
property that
\begin{linenomath}
$$P\in \bigcup_{(T,\phi_T)\in \mathcal M}
\phi_T(S_T\smallsetminus X_T) \text{ implies } P\in
\phi_T(S_T\smallsetminus X_T) \text{ for a unique
$T$}.$$
\end{linenomath}

\item We say that \emph{numerical parameters are generically
locally identifiable on a tree $T$} if there is a proper
algebraic variety $Y_T$ such that for all
$\mathbf s\in S_T\smallsetminus Y_T$, there is a
neighborhood of $\mathbf s$ on which $\phi_T$ is injective.

\item We say the \emph{model $\mathcal M$ is generically locally
identifiable} if the tree parameter is generically identifiable, and
for each tree the numerical parameters are generically locally
identifiable. \end{enumerate}
\end{defn}

Note that the notion of `generic' here is used to mean
`for all parameters but those lying on a proper
subvariety of the parameter space,' and such a variety
is necessarily of lower dimension than the full parameter
space. Using the standard measure
on the parameter space, viewed as a subset of $\R^N$,  this notion thus
also implies `for all
parameters except those in a set of measure 0.'

\smallskip

In the important special case of parameterization maps defined by
polynomial formulas, such as that for the GM+I model, generic local
identifiability of numerical parameters is equivalent to the notion in
algebraic geometry of the map $\phi_T$ being \emph{generically
  finite}. In this case, there exists a proper variety $Y_T$ and an
integer $k$, the degree of the map $\phi_T$, such that restricted to
$S_T\smallsetminus Y_T$ the map $\phi_T$ is not only locally injective
but also $k$-to-1: That is, if $\mathbf s\in S_T\smallsetminus Y_T$
and $P=\phi_T(\mathbf s)$, then the fiber $\phi^{-1}_T(P)$ has
cardinality $k$.

Because of the label swapping issue at internal nodes, for the GM
model and GM+I on an $n$-taxon tree $T$ with vertex set $V$, fibers of
generic points will always have cardinality at least $\kappa!(|V|-n)$.
Thus for these models, the best we can hope for is generic local
identifiability of the model (both tree and numerical parameters)
where the generic fiber has exactly this cardinality.  That in fact is
what we establish in the next section.

\section{Generic Identifiability for the GM+I model}\label{sec:genericId}

We begin our arguments by determining some phylogenetic invariants for
the GM+I model. The notion of a phylogenetic invariant was introduced
by Cavender and Felsenstein \cite{CF87} and Lake \cite{Lake87}, in the
hope that phylogenetic invariants might be useful for practical tree
inference.  Their role here, in proving identifiability, is more
theoretical but illustrates their value in analyzing models.

\smallskip

For a parameterization $\phi_T$ given by polynomial formulas on domain
$S_T\subseteq\R^N$, we may uniquely extend to a polynomial map with
domain $\C^N$, given by the same polynomial formulas, which we again
denote by $ \phi_T: \C^N \longrightarrow {\C}^{\kappa^n}.$

\begin{rem}
  Extending parameters to include complex values is solely for
  mathematical convenience, as algebraic geometry provides the natural
  setting for our viewpoint.  The collection of stochastic joint
  distributions (arising from the original stochastic parameter space)
  is a proper subset of $\im(\phi_T)$.
\end{rem}

The \emph{phylogenetic variety}, $V_T$, is the the smallest algebraic
variety in $\C^{\kappa^n}$ containing $ \phi_T(\C^N)$, \emph{i.e.},
the closure of the image of $\phi_T$ under the Zariski topology,
\begin{linenomath}
  $$V_T=\overline{\im(\phi_T)}\subseteq\C^{\kappa^n}.$$
\end{linenomath}

\begin{rem}
  $V_T$ coincides with the closure of $\im(\phi_T)=\phi_T(\C^N)$ under
  the usual topology on $\C^{\kappa^N}.$ However, while $V_T\cap
  [0,1]^{\kappa^n}$ contains the closure of $\phi_T(S_T)$ under the
  usual topology, these need not be equal.
\end{rem}

Let $\C[P]$ denote the ring of polynomials in the $\kappa^{n}$
indeterminates $\{p_{i_1\dots i_n}\}.$ Then the collection of all
polynomials in $\C[P]$ vanishing on $V_T$ forms a prime ideal $I_T$.
We refer to $I_T$ as a \emph{phylogenetic ideal}, and its elements as
\emph{phylogenetic invariants}.  More explicitly, a polynomial $f\in
\C[P]$ is a phylogenetic invariant if, and only if, $f(P_0)=0$ for
every $P_0\in \phi_T(\C^{\kappa^n})$, or equivalently, if, and only
if, $f(P_0)=0$ for every $P_0\in \phi_T(S_T)$.

\medskip

As we proceed, we consider first the special case of $4$-taxon
trees. We highlight the
$\kappa=2$ case, in part to illustrate the arguments for general
$\kappa$ more clearly, and in part because we can go further in understanding
the 2-state model.

\medskip

\begin{figure}[h]
\begin{center}
\includegraphics[height=.75in]{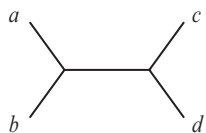}
\end{center}
\caption{The 4-taxon tree $T_{ab|cd}$}\label{fig:4taxa}
\end{figure}

Consider the $4$-taxon binary tree $T_{ab|cd}$, with taxa $a,b,c,d$ as
shown in Figure \ref{fig:4taxa}.

Suppose that $P$ is a $2 \times 2 \times 2
\times 2$  pattern frequency array,
 whose indices correspond to states $[2]=\{1,2\}$
at the taxa in alphabetical order.
Then the internal edge $e$ of $T$ defines the split $ab \mid
cd$ in the tree, and we define the \emph{edge flattening} $F_e$ of $P$
at $e$, a $2^2 \times 2^2$ matrix, by
\begin{linenomath}
\begin{equation}\label{eq:Flat}
F_e =
\begin{pmatrix}
  p_{1111} & p_{1112} & p_{1121} & p_{1122}\\
  p_{1211} & p_{1212} & p_{1221} & p_{1222}\\
  p_{2111} & p_{2112} & p_{2121} & p_{2122}\\
  p_{2211} & p_{2212} & p_{2221} & p_{2222}\\
\end{pmatrix}.
\end{equation}
\end{linenomath}
Notice that the rows of $F_e$ are indexed by the states at $\{ab\}$
and the columns by states at $\{cd\}$. The flattening $F_e$ is
intuitively motivated by considering a `collapsed' model induced by
$e$: taxa $a$ and $b$ are grouped together forming a single variable
$\{ab\}$ with $4$ states, and the grouping $\{cd\}$ forms a second
variable with $4$ states.

This construction can be generalized in a natural way: suppose $T$
is an $n$-taxon tree, and $P$ a $\kappa\times\dots\times\kappa$ array with
indices corresponding to the taxa labeling the leaves of $T$. Then
for any edge $e$ in $T$, we can form from $P$ the matrix $F_{e}$ of size
$\kappa^{n_1} \times \kappa^{n_2}$, where $n_1$ and $n_2$ are the
cardinalities of the two sets of taxa in the split induced by $e$.

From \cite{ARgm} (for a more expository presentation, see also
\cite{ARnme}), we have:

\begin{thm} \label{thm:GM}
  For the $2$-state GM model on a binary $n$-taxon
  tree $T$, the phylogenetic ideal $I_T$ is generated by all $3\times 3$
  minors of all edge flattenings $F_e$ of $P$.
  Moreover, for the $\kappa$-state GM model on an $n$-taxon tree $T$,
  the phylogenetic ideal
  $I_T$ contains all
  $(\kappa+1) \times (\kappa+1)$ minors of all edge flattenings
  of $P$.
\end{thm}

Using this result, we can deduce some
elements of the phylogenetic ideal for
the GM+I model for any number of taxa $n \ge 4$ and any number of
states $\kappa \ge 2$.

\begin{prop}\label{prop:invariants} (Phylogenetic Invariants for GM+I)
\begin{enumerate}
\item \label{prop:inv:item1}
For the $4$-taxon tree $T_{ab|cd}$ and
the $2$-state GM+I model,
the cubic determinantal polynomials
\begin{linenomath}
  $$
  f_1=\left |\begin{matrix}
      p_{1112} & p_{1121} & p_{1122}\\
      p_{1212} & p_{1221} & p_{1222}\\
      p_{2112} & p_{2121} & p_{2122}\\
\end{matrix}\right |
\mbox{ and } f_2=\left |\begin{matrix}
    p_{1211} & p_{1212} & p_{1221}\\
    p_{2111} & p_{2112} & p_{2121}\\
    p_{2211} & p_{2212} & p_{2221}
\end{matrix}\right |
$$
\end{linenomath}
are phylogenetic invariants. These are the two $3\times 3$ minors of
the matrix flattening $F_{ab \mid cd}$ of equation (\ref{eq:Flat}) that do not
involve either of the entries $p_{1111}$ or $p_{2222}$.

\item More generally, for $n\ge 4$ and $\kappa\ge 2$, consider the
  $\kappa$-state GM+I model on an $n$-taxon tree $T$. Then for each
  edge $e$ of $T$, all $(\kappa+1)\times (\kappa+1)$ minors of the
  flattening $F_e$ of $P$ that avoid all entries $p_{ii\dots i}$,
  $i\in[\kappa]$ are phylogenetic invariants.
\end{enumerate}
\end{prop}

\begin{pf}  We prove the first statement in detail.
From equation (\ref{eq:decomp}),
for
any $P=\phi_T(s)$ we have $P=(1-\delta)P_{GM}+\delta
 P_I$, where $P_{GM}$ is a 4-dimensional table arising from the GM
  model on $T$ and $P_I=\diag(\pi_I)$ is a diagonal table with entries
  giving the distribution of states for the invariable sites.
  Flattening these tables with respect to the internal edge of the
  tree, we obtain
\begin{linenomath}
\begin{align}F_{ab \mid cd} &= (1-\delta) F_{GM} + \delta F_I\notag \\
  &=(1 - \delta)
\begin{pmatrix}
  \tilde p_{1111} & \tilde p_{1112} & \tilde p_{1121} & \tilde p_{1122}\\
  \tilde p_{1211} & \tilde p_{1212} & \tilde p_{1221} & \tilde p_{1222}\\
  \tilde p_{2111} & \tilde p_{2112} & \tilde p_{2121} & \tilde p_{2122}\\
  \tilde p_{2211} & \tilde p_{2212} & \tilde p_{2221} & \tilde
  p_{2222}
\end{pmatrix}
+
\delta \begin{pmatrix}
  \pi_I(1) &\ 0\ &\  0\ & 0 \\
  0 & 0 & 0 & 0\\
  0 & 0 & 0 & 0\\
  0 & 0 & 0 & \pi_I(2) \\
\end{pmatrix}.\label{eq:Psum}
\end{align}
\end{linenomath}
By Theorem \ref{thm:GM}, all $3\times 3$ minors of $F_{GM}$ vanish.
Since the `upper right' and `lower left' minors of $F_{ab|cd}$ are the
same as those of $F_{GM}$, up to a factor of $(1-\delta)^3$, they also
vanish.

Straightforward
modifications to this argument give the general case.\hfill\qed
\end{pf}

For arbitrary $n,\kappa$, the GM+I model should have many other
invariants than those found here.
Among these is, of course, the stochastic invariant
\begin{linenomath}
$$f_s(P)=1-\sum_{\mathbf i\in [\kappa]^n} p_{\mathbf i}.$$
\end{linenomath}

In the simplest interesting case of the GM+I model, however, we
have the following computational result.

\begin{prop}\label{prop:invariantsK2} The phylogenetic ideal
  for the $2$-state GM+I model on the $4$-taxon tree $T_{ab|cd}$ of Figure
  \ref{fig:4taxa} is generated by $f_s$ and
the minors $f_1$, $f_2$ above;
\begin{linenomath}
$$I_T  = \langle f_s, f_1, f_2 \rangle.$$
\end{linenomath}
\end{prop}

\begin{pf}
  A computation of the Jacobian of the parameterization
$\phi_T: S \subset \C^{13} \to \C^{2^4}$ shows it has full rank at
some points, and so $V_T$ is of dimension 13.  If $I =
\langle f_s,f_1, f_2 \rangle$, then $I \subseteq I_T$.  Another computation
shows that $I$ is prime and of
  dimension $13$.  Thus, necessarily $I =
  I_T$.  (The code for these computations is given in Appendix
  \ref{app:code}.)\hfill\qed
\end{pf}

Let $V_{ab | cd}$, $V_{ac | bc}$, $V_{ad | bc}$ be the varieties for
the $2$-state GM+I models for the three $4$-taxon binary tree topologies, with
corresponding phylogenetic ideals $I_{ab \mid cd}$, $I_{ac \mid bd}$,
$I_{ad \mid bc}$.
Of course Proposition \ref{prop:invariantsK2}
gives generators for each of these ideals --- two $3
\times 3$ minors of the flattenings of $P$ appropriate to those tree
topologies, along with $f_s$. A computation (see Appendix \ref{app:code}) shows
that these three ideals are distinct. Therefore the three varieties
are distinct, and their pairwise intersections are proper
subvarieties. Thus for any parameters $\mathbf s$
not lying in the inverse image
of these subvarieties, $T$ is uniquely determined from $\phi_T(\mathbf s)$.
Thus we obtain

\begin{cor} \label{cor:identTreeN4k2}
  For the $2$-state GM+I model on binary 4-taxon trees, the tree parameter is
  generically identifiable.
\end{cor}

As $\dim(V_{ab|cd})=13$, and the parameter space for $\phi_T$ is 13
dimensional, we also immediately obtain that the map $\phi_T$ is
generically finite. This yields

\begin{cor} \label{cor:idenNumN4k2}
For the $2$-state GM+I model on a binary 4-taxon tree,
numerical parameters are generically locally
identifiable.
\end{cor}

Note that this does approach does not yield the cardinality of the
generic fiber of the parameterization map, which is also of
interest. We will return to this issue in Theorem
\ref{thm:genericIdent}.

\medskip

Further computations show that
$\dim(V_{ab|cd} \cap V_{ac|bd}\cap V_{ad|bc})=11$. As this
intersection contains all points arising from the GM+I
model on the 4-taxon star tree, which is an 11-parameter model, this
is not surprising. In fact, one can verify computationally
that the ideal $I_{ab|cd}+I_{ac|bd}+I_{ad|bc}$ is the defining
prime ideal of the star-tree variety.
We also note that the ideal $I_{ab|cd} + I_{ac|bd}$
decomposes into two primes, both of dimension 11. Thus the variety
defined by this ideal has two components, one of which is the variety
for the star tree.

\medskip

In principle, the ideal $I_T$ of all invariants for the GM+I model
on an arbitrary tree $T$ can be computed from the parameterization
map $\phi_T$ via an elimination of variables using Gr\"obner bases
\cite{MR2001c:92009}. However, if all invariants for the
$\kappa$-state GM model on $T$ are known, they can provide an
alternate approach to finding $I_T$ which, while still proceeding by
elimination, should be less computationally demanding.

To present this most simply, we note that because
our varieties lie in the hyperplane described by the stochastic invariant,
it is natural to consider their projectivizations,
lying in $\mathbb P^{\kappa^n-1}$ rather than $\C^{\kappa^n}$. The
corresponding phylogenetic ideals, which we denote by $J_T$,
are generated by the homogeneous polynomials in $I_T$, and do not contain the
stochastic invariant. Conversely, $I_T$ is generated by the elements of
$J_T$ together with the stochastic invariant.

In addition, we need
not restrict ourselves to the GM model, but rather deal with any
phylogenetic model parameterized by polynomials.

\begin{prop} \label{prop:elim}
Suppose $\widetilde \phi_T:\C^N\to \C^{\kappa^n}$ is a
parameterization map for some phylogenetic model $\mathcal M$ on
$T$, with corresponding homogeneous phylogenetic ideal $\widetilde
J_T$. Let
\begin{linenomath}
$$\phi_T:\C^{N}\times\C^\kappa\to
 \C^{\kappa^n}$$
\end{linenomath}
be the parametrization map for the $\mathcal M$+I model
given by
\begin{linenomath}
$$\phi_T(\mathbf s,(\delta,\boldsymbol \pi_I))=(1-\delta)
\widetilde \phi_T(\mathbf s)+\delta \diag(\boldsymbol \pi_I).$$
\end{linenomath}
Let
$P'$ denote the collection of all indeterminate entries of $P$
except those in $P_{eq}=\{p_{ii\dots i}\mid i\in[\kappa]\}$. Then
the homogeneous phylogenetic ideal $J_T$ for the $\mathcal M$+I
model on $T$ is $J_T=\left (\widetilde J_T\cap \C[P'] \right)\C[P].$
Thus $J_T$ can be computed from $\widetilde J_T$ by elimination of
the variables in $P_{eq}$.
\end{prop}

\begin{pf}
Extend the parameterization maps $\widetilde \phi_T, \phi_T$ to
parameterizations of cones by introducing an additional parameter,
\begin{linenomath}
$$ \widetilde \Phi_T(\mathbf s,t)=t\,\widetilde\phi_T(\mathbf s)$$
$$
\Phi_T(\mathbf s,(\delta,\boldsymbol \pi_I ),t)
=t\,\phi_T(\mathbf s,(\delta,\boldsymbol \pi_I))$$
\end{linenomath}
Then
$\im(\Phi_T)=\C^\kappa\times\operatorname{proj}(\im(\widetilde \Phi_T)),$
where $\C^\kappa$ corresponds to coordinates in $P_{eq}$ and
`$\operatorname{proj}$' denotes
the projection map from $P$-coordinates to $P'$-coordinates. As $J_T$
is the ideal of polynomials vanishing on $\im(\Phi_T)$, and
$\tilde J_T\cap \C[P']$ the ideal
vanishing on $\operatorname{proj}(\im(\widetilde \Phi_T))$, the result follows.
\hfill\qed
\end{pf}

Using this, in the appendix we give an alternate computation to show
both part (\ref{prop:inv:item1}) of Proposition
\ref{prop:invariants}, and Proposition \ref{prop:invariantsK2}.
While this computation is quite fast, a more naive attempt to find
GM+I invariants directly from the full parameterization map using
elimination was unsuccessful, demonstrating the utility of the
proposition.
Moreover, we can use this proposition to compute all 2-state GM+I invariants
on the 5-taxon binary tree as well. This leads us to

\begin{conj} On an $n$-taxon binary tree, the ideal of homogeneous
invariants for the 2-state
GM+I model is generated by those $3\times 3$
minors of edge flattenings
that do not involve the variables $p_{11\dots1}$ and $p_{22\dots2}$,
together with the
stochastic invariant.
\end{conj}

\medskip

Although we are unable to determine all GM+I invariants for the
4-taxon tree for general $\kappa$, using only those described in
Proposition \ref{prop:invariants} we can still obtain
identifiability results through a modified argument.

\begin{prop}\label{prop:treeId} For the $\kappa$-state GM+I model on
binary 4-taxon trees, $\kappa\ge 2$, the tree parameter is
generically identifiable.
\end{prop}

\begin{pf} By the argument leading to Corollary \ref{cor:identTreeN4k2},
it is enough to show the varieties $V_{ab \mid cd}$, $V_{ac
    \mid bd}$, and $V_{ad|bc}$ are distinct.
Considering, for example, the first two, we can
show that the varieties $V_{ab \mid cd}$ and $V_{ac
    \mid bd}$ are distinct, by giving an invariant $f \in I_{ac \mid
    bd}$  and a point $P_0\in V_{ab|cd}$
such that $f(P_0)\ne 0$.

Using Proposition \ref{prop:invariants}, we pick an
  invariant $f \in I_{ac \mid bd}$ as follows: In the flattening
$F_{ac|bd}$ according to the split $ac|bd$, choose any collection
of $\kappa+1$ $ac$-indices with distinct $a$ and $c$ states, \emph{e.g.},
$\{12,13,\dots,1\kappa,21,23\}$. Using the same set as $bd$-indices,
this determines a $(\kappa+1)\times(\kappa+1)$-minor $f$.

We pick $P_0=\phi_{T_{ab|cd}}(\mathbf s)$ using the parameterization
of equation (\ref{eq:Pdef}) by making a specific choice of parameters
$\mathbf s$. On $T_{ab|cd}$, with the root $r$ located at one of the
internal nodes, choose parameters $\mathbf{s}$ as follows: Let
$\pr$, $\pI$ be arbitrary but with all entries of $\pr$ positive.
Pick any $\delta \in [0,1)$. For the four terminal edges choose
$M_e$ to be the $\kappa \time \kappa \times \kappa$ identity matrix
$I_\kappa$. For the single internal edge $e$ of $T$, choose any
Markov matrix $M_{e}$ with all positive entries. For such
parameters, the entries of the joint distribution $P_0 =
\phi_{T_{ab|cd}} (\mathbf{s})$ are zero except for the pattern
frequencies $p_{iijj}$, where the states at the leaves $a$ and $b$
agree and the states at the leaves $c$ and $d$ agree.  Since the
entries of $M_{e}$ and the root distributions are positive, each of
the $p_{iijj} > 0$.

But considering the flattening $F_{ac \mid bd}$ of
$P_0=\phi_{T_{ab|cd}} (\mathbf{s})$ with respect to the `wrong'
topology $T_{ac \mid bd}$, we observe that the $\kappa^2$ non-zero
entries $p_{iijj}$ of $F_{ac \mid bd}$ all lie on the diagonal of
$F_{ac \mid bd}$, in the positions with $ij$ as both $ac$-index and
$bd$-index. Furthermore, by our choice of $f$, a subset of them
forms the diagonal of the submatrix whose determinant is $f$.
Therefore $f(P_0)\ne 0$.\hfill\qed
\end{pf}

\begin{prop}(Recovery of invariable site parameters)\label{prop:idformulas}
\begin{enumerate}
\item For the 4-taxon tree $T_{ab|cd}$ and the 2-state GM+I model, suppose
$P=\phi_T(\mathbf s)$. Then generically the parameters in
$\mathbf s$ related to invariable sites can be recovered from $P$ by
the following formulas:
\begin{linenomath}
$$\delta=\frac {|A_1|+|A_2|}{|B|},\ \  \boldsymbol \pi_I=\frac 1{|A_1|+|A_2|} \left (
|A_1|,|A_2|\right ),$$ where $B=\begin{pmatrix}
    p_{1212} & p_{1221} \\
    p_{2112} & p_{2121}
\end{pmatrix}$,
$$A_1=\begin{pmatrix}
      p_{1111} & p_{1112} & p_{1121}\\
      p_{1211} & p_{1212} & p_{1221}\\
      p_{2111} & p_{2112} & p_{2121}\\
\end{pmatrix}, \ \
A_2=\begin{pmatrix}
    p_{1212} & p_{1221} & p_{1222}\\
    p_{2112} & p_{2121} & p_{2122}\\
    p_{2212} & p_{2221} & p_{2222}
\end{pmatrix}.
$$
\end{linenomath}
\item More generally, for the $\kappa$-state GM+I model on $T_{ab|cd}$,
the invariable site parameters can be recovered
from a generic point in the image of the parameterization map by
rational formulas of the form
\begin{linenomath}
$$\delta=\frac
{\sum_{i\in[\kappa]}|A_i|}{|B|}, \ \ \boldsymbol \pi_I=\frac
1{\sum_{i\in[\kappa]} |A_i|} \left ( |A_1|,|A_2|,\dots, |A_n|\right
).$$
\end{linenomath}
Here $|B|$ is any $\kappa \times \kappa$ minor of $F_{ab|cd}$
that omits the all rows and columns indexed by $ii$, and $|A_i|$ is
the $(\kappa+1)\times(\kappa+1)$ minor obtained by including all
rows and columns chosen for $B$ and in addition the $ii$ row and
$ii$ column.
\end{enumerate}
\end{prop}

\begin{pf} We
  give the complete argument in the case $\kappa = 2$ first.  For a
  joint distribution $P \in \im(\phi_T)$, write $F_{ab \mid cd} =
  (1-\delta) F_{GM} + \delta F_I$ as in equation (\ref{eq:Psum}).  Since
$A_1$ is the `upper left'  $3 \times 3$ submatrix of $F_{ab \mid
cd}$, using linearity properties of the determinant, and that all $3
\times 3$ minors of $F_{GM}$ evaluate to zero, we observe that
\begin{linenomath}
\begin{align*}
  \vert A_1 \vert
&=(1 - \delta)^3 \left|
\begin{matrix}
  \tilde p_{1111} & \tilde p_{1112} & \tilde p_{1121} \\
  \tilde p_{1211} & \tilde p_{1212} & \tilde p_{1221} \\
  \tilde p_{2111} & \tilde p_{2112} & \tilde p_{2121} \\
\end{matrix}
\right| + \left|
\begin{matrix}
  \delta \pi_I(1) & 0 &\ 0 \\
  0 & (1-\delta) \tilde p_{1212} &\ (1-\delta) \tilde p_{1221} \\
  0 & (1-\delta) \tilde p_{2112} &\ (1-\delta) \tilde p_{2121} \\
\end{matrix}\right|\\
\\
&= \delta \pi_I(1) \left|
\begin{matrix}
  (1-\delta)\tilde p_{1212} &\  (1-\delta)\tilde p_{1221} \\
 (1-\delta) \tilde p_{2112} &\ (1-\delta )\tilde p_{2121} \\
\end{matrix}\right|.
\end{align*}
\end{linenomath}
Thus we have $\vert A_1 \vert = \delta \pi_I(1) \vert B \vert$. Now,
if $\vert B \vert \neq 0$, then
\begin{linenomath}
$$
\delta \pi_I(1) = \frac{\vert A_1 \vert}{\vert B \vert}.
$$
\end{linenomath}
As $|B|$ does not vanish on all of $V_T$, we have a rational formula
to compute $\delta \pi_I(1)$ for generic points on $V_T$.

Similarly, since $A_2$ is the `lower right' submatrix of $F_{ab \mid
cd}$, then
\begin{linenomath}
$$\delta \pi_I(2) =
\frac{\vert A_2 \vert}{\vert B \vert}.
$$
\end{linenomath}
Adding these together, we obtain the stated rational expression for
$\delta$.

Assuming additionally the generic condition that $\delta \neq 0$, then we find
\begin{linenomath}
$$\boldsymbol \pi_I =\left ( \frac{\vert A_1
  \vert}{\vert A_1\vert + \vert A_2 \vert},
\frac{\vert A_2
  \vert}{\vert A_1 \vert + \vert A_2 \vert}\right ).$$
\end{linenomath}
Thus the parameters $\delta, \boldsymbol \pi_I$ are
generically identifiable for GM+I on $T$.

One readily sees the argument above can be modified for arbitrary
$\kappa$.\hfill\qed
\end{pf}

Note that when $\kappa>2$ the above proposition gives many
alternative rational formulas for the invariable site parameters, as
there are many options for choosing the matrix $B$.

We now obtain our main result.

\begin{thm}\label{thm:genericIdent}
The $\kappa$-state GM+I model on $n$-taxon binary trees, with $n\ge
4$, $\kappa \ge 2$, is generically locally identifiable.
Furthermore, for an $n$-taxon tree with $V$ vertices, the fibers of
generic points of $V_T$ under the parametrization map have
cardinality $\kappa!(|V|-n)$. Thus for generic points, label
swapping at internal nodes is the only source of
non-identifiability.
\end{thm}

\begin{pf} Suppose $T$ is an $n$-taxon tree with $P=\phi_T(\mathbf
s)$. Choose some subset of 4 taxa, say $\{a,b,c,d\}$, and suppose
the induced quartet tree is $T_{ab|cd}$. Then $P_{abcd}$, the
4-marginalization of $P$, is easily seen to be of the form
$P_{abcd}=\phi_{T_{ab|cd}}(\mathbf s_{abcd})$ where $\mathbf
s_{abcd}=g(\mathbf s)$ and $g$ is a surjective polynomial function.
But the tree
$T_{ab|cd}$ is generically identifiable by Proposition
\ref{prop:treeId}, and thus invariable site parameters in $\mathbf s_{abcd}$
are generically identifiable by Proposition \ref{prop:idformulas}.
As these coincide with the invariable site parameters in $\mathbf
s$, and generic conditions on $\mathbf s_{abcd}$ imply generic
conditions on $\mathbf s$, the invariable site parameters are
generically identifiable for the full $n$-taxon model.

As an $n$-taxon binary tree topology is determined
by the collection of all induced quartet tree topologies, one can now see
that $T$ is generically identifiable. Alternately,
using the identified invariable site parameters,
and assuming the additional
generic condition that $\delta\ne 1$, note that
\begin{linenomath}
$$P_{GM} =
\frac{1}{(1-\delta)} \left( P - \delta P_I\right)
$$
\end{linenomath}
is a joint
distribution arising from general Markov parameters. Thus generic
identifiability of the tree can also by obtained from
Steel's  result for the GM model \cite{S94} applied to $P_{GM}$.

The generic identifiability of the remaining numerical parameters follows
from Chang's argument \cite{MR97k:92011} applied to $P_{GM}$.
Chang's approach also indicates the cardinality of the generic fiber is
$\kappa!(|V|-n)$ due to the label swapping phenomenon.\hfill\qed
\end{pf}

\section{Estimating Invariable Sites Parameters}\label{sec:estInv}

The concrete result in Proposition \ref{prop:idformulas} gives
explicit rational formulas for recovering parameters relating to
invariable sites from the joint distribution. These can be viewed as
generalizations of the formulas found in \cite{SHL00} for
group-based models. As \cite{SHL00} develops the group-based model
formulas into a heuristic means of estimating the invariable site
parameters from data without performing a full Maximum Likelihood
fit of data to a tree under a $\mathcal M$+I model, one might
suspect the formulas of Proposition \ref{prop:idformulas} could be
used similarly without the need to assume $\mathcal M$ was
group-based, or approximately group-based.
 We emphasize that however useful such an
estimate might be, it would not be intended to replace a more
statistical but time-consuming computation, such as obtaining the
Maximum Likelihood estimates for these parameters.

However, it is by no means obvious how to use these formulas well
even for a heuristic estimate. First, for a 4-taxon tree
we have many choices for the
matrix $B$, in fact
\begin{linenomath}
$$\binom{\kappa^2-\kappa}{\kappa}^2$$
\end{linenomath}
of them, so even for $\kappa=4$, there are 245,025 basic sets of the
formulae. Moreover, while these simple formulae
emerged from our method of proof, one could in fact modify them by
adding to any of them a rational function whose numerator is a
phylogenetic invariant for the GM+I model, and whose denominator is
not. Since the invariant vanishes on any joint distribution arising
from the model, the resulting formulae will still recover invariable
site information for generic parameters. Thus there are actually
infinitely many formulas for recovering invariable site parameters.

One can nonetheless consider simple averaging schemes using only the
basic formulas of Proposition \ref{prop:idformulas} and find that on
simulated data they perform quite well at approximately recovering
invariable site parameters from empirical distributions. However,
averaging the large number of formulas give here, and then also
averaging over a large sample of quartets,
 as is proposed in \cite{SHL00}, is more
time consuming than one might wish for a fast heuristic. Moreover,
one must be aware that the denominator in these formulas may vanish
on an empirical distribution --- it is certain to be non-zero only
for true distributions for GM+I arising from generic parameters.

Nonetheless, it would be of interest to develop versions of these
formulas with good statistical estimation properties, as the GM+I
model encompasses models such as the GTR+I model which is often
preferred in biological data analysis to group-based+I models. Of
course addressing more general rate-variation models would be even
more desirable, though our results here are not sufficient for that.

\appendix

\section{Code for Computational Algebra Software}\label{app:code}

The following code is also available on the authors' websites.

\subsection{Computation for Proposition \ref{prop:invariantsK2} }

To show the variety has dimension 13, we execute the following Maple code:

{
\scriptsize
\begin{verbatim}
pa := Matrix([[p,1-p]]); Mae := Matrix([[1-a,a],[r,1-r]]);
Meb := Matrix([[1-b,b],[s,1-s]]); Mef := Matrix([[1-e,e],[t,1-t]]);
Mfc := Matrix([[1-c,c],[u,1-u]]); Mfd := Matrix([[1-d,d],[v,1-v]]);
P := Array(1..2,1..2,1..2,1..2);
for i from 1 to 2 do for j from 1 to 2 do for k from 1 to 2 do for l from 1 to 2 do
  P[i,j,k,l]:=0;
  for m from 1 to 2 do  for n from 1 to 2 do
    P[i,j,k,l]:=P[i,j,k,l]+pa[1,i]*Mae[i,m]*Meb[m,j]*Mef[m,n]*Mfc[n,k]*Mfd[n,l];
  od;od;
  P[i,j,k,l]:=(1-w)*P[i,j,k,l];
od;od;od;od;
P[1,1,1,1]:=P[1,1,1,1]+w*q: P[2,2,2,2]:=P[2,2,2,2]+w*(1-q):
Q:=ListTools[Flatten](convert(P,listlist)):
J:=VectorCalculus[Jacobian](Q,[a,b,c,d,e,r,s,t,u,v,p,q,w]):
K:=subs({a=1/3,b=1/5,c=1/7,d=1/11,e=1/13,r=1/17,s=1/19,t=1/23,u=1/29,v=1/31,
                                                          p=1/3,q=1/5,w=1/7},J):
LinearAlgebra[Rank](K);
\end{verbatim}
}

Using Singular \cite{sing}, we complete the proof:

{
\scriptsize
\begin{verbatim}
LIB "matrix.lib";  LIB "primdec.lib";
ring r = 0, (p0,p1,p2,p3,p4,p5,p6,p7,p8,p9,p10,p11,p12,p13,p14,p15),dp;
// Define matrix flattening F_{ab | cd} and polys fs, f1, f2
matrix Fab[4][4]=p0,p1,p2,p3,p4,p5,p6,p7,p8,p9,p10,p11,p12,p13,p14,p15;
matrix UR[3][3]=submat(Fab,1..3,2..4); matrix LL[3][3]=submat(Fab,2..4,1..3);
poly f1=det(UR); poly f2=det(LL);
poly fs = p0+p1+p2+p3+p4+p5+p6+p7+p8+p9+p10+p11+p12+p13+p14+p15-1;
ideal I = fs,f1,f2;   // define ideal I
dim(std(I));          // compute dimension of r/I
primdecGTZ(I);        // compute primary decomposition of I to show prime
\end{verbatim}
}

\subsection{Computation for intersections of $V_{ab|cd},V_{ac|bd},V_{ad|bc}$}

Continuing the Singular session above, we execute the following:

{
\scriptsize
\begin{verbatim}
/* Define ideals Iac, Iad corresponding to two alternative tree
   topologies for 4-taxon trees.  (So, I = Iab in this notation.)   */
// Flattening for ac | bd split
matrix Fac[4][4]=p0,p1,p4,p5,p2,p3,p6,p7,p8,p9,p12,p13,p10,p11,p14,p15;
poly f3=det(submat(Fac,1..3,2..4)); poly f4=det(submat(Fac,2..4,1..3));
ideal Iac = fs,f3,f4;
// Flattening for  ad | bc split
matrix Fad[4][4]=p0,p2,p4,p6,p1,p3,p5,p7,p8,p10,p12,p14,p9,p11,p13,p15;
poly f5=det(submat(Fad,1..3,2..4)); poly f6=det(submat(Fad,2..4,1..3));
ideal Iad = fs,f5,f6;
reduce(f1,std(Iac));  // non-zero answer shows f1 not in Iac
reduce(Iac,std(I));   // non-zero shows f3,f4 not in I
ideal J = I,Iac; dim(std(J));  // show dim is 11
ideal K = J,Iad; dim(std(K));  // show dim is 11
primdecGTZ(K);        // show K prime, and thus ideal for star tree
\end{verbatim}
}

\subsection{Computation of 2-state GM+I ideal, 4-taxon trees, using Proposition \ref{prop:elim} }

The following Singular code performs the needed elimination for a binary tree:

{
\scriptsize
\begin{verbatim}
ideal Igm = minor(Fab,3);
// Eliminate the `diagonal' variables
ideal Igmi = elim1(Igm,p0*p15);
\end{verbatim}
}

For the star tree, the 2-state GM ideal is known from \cite{ARgm}.
Thus elimination can be used to find GM+I invariants. We also show
this result agrees with $\mathtt K$ above.

{
\scriptsize
\begin{verbatim}
ideal Igm = minor(Fab,3),minor(Fac,3),minor(Fad,3);
// Eliminate the `diagonal' variables
ideal Igmi = elim1(Igm,p0*p15),fs;
reduce(K,std(Igmi));  // all 0's indicates ideal containment
reduce(Igmi,std(K));  // all 0's indicates ideal containment
\end{verbatim}
}

\bibliographystyle{elsart-num} \bibliography{Phylo}

\end{document}